\documentclass[%
 reprint,
 amsmath,amssymb,
 aps,
pra,
]{revtex4-1}

\usepackage{graphicx}
\usepackage{dcolumn}
\usepackage{bm}

\begin{document}

\preprint{APS/123-QED}

\title{Generation of non-local evolution loops and exchange operations for quantum control in three dimensional anisotropic Ising model}

\author{Francisco Delgado}
\altaffiliation[Also at ]{Escuela Nacional de Posgrado en Ciencias e Ingenier\'ia, Tecnol\'ogico de Monterrey, M\'exico.}
\email{fdelgado@itesm.mx}
\affiliation{Departamento de F\'isica y Matem\'aticas, Tecnol\'ogico de Monterrey, Campus Estado de M\'exico, Atizap\'an, Estado de M\'exico, CP. 52926, M\'exico.}%

\date{\today}

\begin{abstract}
Control of quantum entanglement has been considered as elemental physical resource for quantum applications in Quantum Information and Quantum Computation. Control of entangled states on a couple of atoms, ions or quantum dots are milestones in almost all quantum applications towards a scalable spin-based quantum computers or quantum devices. For magnetic systems, Ising model is an interaction which generates and modifies entanglement properties of quantum systems based on matter. In addition, when this interaction includes driven magnetic fields, it can be controlled to sustain, characterize or modify entanglement and other quantum properties. In this work, recent results about evolution in a general anisotropic three dimensional Ising model including an inhomogeneous magnetic field is considered to obtain some general quantum control effects for their sustainability, programmed evolution or transformation: Evolution loops and Exchange operations. This control is achievable through a set of physical parameters, whose prescriptions are reported. The use of a non local basis in the model to express time evolution lets take advantage to describe and control the system, in particular with those issues associated with entanglement and operations mentioned before. Finally, some analysis about equivalent gates based on our development is made including an example with teleportation, using one of the gates constructed.

\begin{description}
\item[PACS numbers]
03.67.-a; 03.67.Bg; 42.50.Dv; 03.65.Ud; 03.67.Hk; 03.65.Aa
\end{description}
\end{abstract}

\maketitle


\section{Introduction}

Control of quantum systems has been developed in several directions. Since control related with a fine management of physical variables until control based on specific interactions, in order to exploit them for concrete purposes. Quantum entanglement has been exploited by quantum applications as a central aspect to improve information processing in terms of capacity and speed \cite{jozsa1, jozsa2, bennet1}. Codification and management of information in quantum terms is an alternative improved method in contrast with classical trends. Since fundamental applied proposals about Quantum Computation \cite{feyn1, deutsch1, steane1} and Quantum Cryptography \cite{bennet2, ekert1}, quantum control has been developed as research area. For this reason, research around of entanglement control, their complexity, properties and potential usefulness \cite{bennet5, nielsen1} is a basic aspect in Quantum Mechanics \cite{horo1} necessarily associated with these applied developments.

Entanglement will not have a complete map of road until its quantification and behavior could be understood since quantum interactions which generates it. In the last sense, in nowadays, Hamiltonian models, which are able to generate entanglement, are actually being studied in order to understand how this quantum feature is generated on several physical systems. For magnetic systems, Ising model \cite{ising1,brush1} in quantum mechanics terms is a Hamiltonian model derived from interaction between spin systems and potentially driven by external magnetic fields. Nielsen work \cite{nielsen2}, reported by first time how spin systems using this model can generate and expose entanglement evolution. Thus, another aim of entanglement research is its control in several terms: generation, sustainability and maximization. Development of quantum control has been boosted by notable works in nanoscale systems \cite{wise1}, quantum feedback control \cite{dohe1,dohe2} and in quantum systems under continuous feedback \cite{wise2,wise3,clark1}. 

Ising interaction has been studied in terms of transference and control of entanglement in bipartite qubits \cite{terzis1}, chains or lattices \cite{niem1,pfeuty1,stelma1,novo1}. After of these approaches, different studies have extended this research for more complex systems and depending on external parameters (temperature, strength of external field, geometry, etc.) \cite{arnesen1, wang1, khaneja1}. Nowadays, possibility to control quantum states of a single or a couple of electron spins, in particular in quantum dots or electronic gases, is still at the heart of developments towards a scalable spin-based quantum computer because this control in combination with controlled exchange between neighboring spins, would let obtain universal quantum operations \cite{recher1, saraga1, kopp1} in agreement with DiVincenzo criteria \cite{vincenzo1} in terms of reliability of state preparation and identification of well identified qubits. 

The aim of this paper is develop control schemes based on Evolution Loops and Exchange Operations on a magnetic bipartite system ruled by a general three dimensional anisotropic Ising interaction including an inhomogeneous magnetic field in a several fixed directions one at time. This model includes several models as particular cases, which have been considered in those previous works. In this approach, analysis of dynamics is conducted on a non-local basis in terms of classical Bell states in order to remark algebraic aspects found for this interaction around entanglement, obtaining direct applications about its control in terms of sustainability and manipulation.

\section{Driven anisotropic Ising Hamiltonian in three dimensions and time evolution in a non local basis}

Control in Ising interaction (for different models: XX, XY, XYZ, etc.) has been analyzed several systems and configurations \cite{berman1,wang2,aless1}. In analysis about structured quantum control effects \cite{branczyk1, xi1, delgado1, delgado2}, different versions of Ising interaction are considered in terms of physical elements or configurations: homogeneity of magnetic fields, dimensions, restrictions in dimensions, number of particles involved and strength of external fields used. Normally, because these aspects generate simplifications on geometry as in properties of physical systems involved too \cite{wang2, kamta1, sun1, zhou1, gunlycke1}. 

To develop results related with control, we adopt the following Hamiltonian recently developed for the bipartite anisotropic Ising model \cite{delgadoA} including an inhomogeneous magnetic field restricted to the $h$-direction ($h=1,2,3$ at time, corresponding with $x,y,z$):

\begin{eqnarray} \label{hamiltonian}
H_h&=&-\mathbf{\sigma_1 \cdot J \cdot \sigma_2}+\mathbf{B_1 \cdot \sigma_1}+\mathbf{B_2 \cdot \sigma_2} \nonumber \\
&=& -\sum_{k=1}^3 J_k {\sigma_1}_k {\sigma_2}_k+{B_1}_h {\sigma_1}_h+{B_2}_h {\sigma_2}_h\ 
\end{eqnarray}

\noindent it generalizes several models considered in control in the previously cited works. In the same terms and notation that \cite{delgadoA}, diagonalization of Hamiltonian to obtain the corresponding eigenvalues which are independent of $h$:

\begin{eqnarray} \label{eigenvalues}
{\mathcal{E}_h}^{(1)}=-J_h-{R_h}_+ &,& 
{\mathcal{E}_h}^{(2)}=-J_h+{R_h}_+ \\ \nonumber
{\mathcal{E}_h}^{(3)}= \quad J_h-{R_h}_- &,& 
{\mathcal{E}_h}^{(4)}= \quad J_h+{R_h}_- 
\end{eqnarray}

\noindent where ${R_h}_-$ and ${R_h}_+$ are defined as:

\begin{eqnarray} \label{erres}
{R_h}_\pm &=&\sqrt{{B_h^2}_\pm+{J_{i,j}^2}_\mp}=\sqrt{{B_h^2}_\pm+{J_{\{ h \}}^2}_\mp}
\\ \nonumber
{\rm with:} && J_{\{ h \}\pm} \equiv {J_{i,j}}_\pm=J_i \pm J_j \\ \nonumber 
&& B_{h \pm}=B_{1_h}\pm B_{2_h}
\end{eqnarray}

\noindent and $h,i,j$ is understood as a cyclic permutation of $1,2,3$ which it is simplified in that work by using $\{ h \}$ as equivalent to the pair of scripts $i,j$. Note that $U(t) \in SU(4)$ because the sum of eigenvalues is zero.

\subsection{Definitions and notation}

Using the original and practical notation used in \cite{delgadoA}, we set:

\begin{eqnarray} \label{defs}
{b_h}_\pm=\frac{{B_h}_\pm}{{R_h}_\pm} &,& {j_h}_\pm=\frac{{J_{\{h\}}}_\mp}{{R_h}_\pm} \in [-1,1]
\end{eqnarray}

As is remarked there, subscripts $-,+$ are settled for these physical variables remarking their internal operations. When Bell basis is used to set the evolution operator as privileged basis, it is convenient introduce several changes in the custom notation by using $-,+$ as some lower and upper scripts, which could evolve to $-1,+1$ if they appear in mathematical expressions. Following with the notation settled in that work, capital scripts $A, B, ...$ are reserved for $0, 1$ referred to the computational basis; greek scripts for $-1,+1$ or $-,+$; latin scripts $h,i,j,k,...$ for spatial coordinates $x,y,z$ or $1,2,3$; and $a,b,c,...$ (between parenthesis) as subscripts to denote energy levels $1,2,3,4$ when will be required. $\cdot$ is used sometimes to emphasize number multiplication between terms in scripts and avoid confusions. Thus, in this notation the standard Bell states are:

\begin{eqnarray} \label{bellnotation}
\left| \beta_{--} \right> \equiv \left| \beta_{00} \right> &,& 
\left| \beta_{-+} \right> \equiv \left| \beta_{01} \right> \\ \nonumber 
\left| \beta_{+-} \right> \equiv \left| \beta_{10} \right> &,&  
\left| \beta_{++} \right> \equiv \left| \beta_{11} \right>    
\end{eqnarray}

The energies $\mathcal{E}_h^{(a)}$ corresponding with $E_{\mu \nu}: E_{--},E_{-+},E_{+-},E_{++}$ in this notation, where:

\begin{eqnarray} \label{eigenvalues2}
{E_h}_{\mu \nu}&\equiv&{\mathcal{E}_h}^{(2+\mu+\frac{1+\nu}{2})}= \mu J_h+\nu {R_h}_{-\mu}  \\ \nonumber
& =&\mu J_h+\nu \sqrt{{B_h}^2_{-\mu}+{J^2_{\{h\}}}_{\mu}}
\end{eqnarray}

\noindent for each direction $h$. Their corresponding eigenstates are:

\begin{align} \label{eigenvectors}
\left| \phi_{\mu \nu}^{1} \right> =&\sum_ {\epsilon \in \{-,+\}} \frac {\delta_{+\epsilon}\nu (1+\mu \nu {j_1}_{-\mu})-\delta_{- \epsilon}\mu {b_1}_{-\mu} }{\sqrt{2}\sqrt{1+\nu\mu {j_1}_{-\mu}}} \left| \beta_{\mu \epsilon} \right> \nonumber \\ \nonumber 
\left| \phi_{\mu \nu}^{2} \right> =&\sum_{\epsilon \in \{-,+\}} \frac {\delta_{+\epsilon}i \nu (1+\mu \nu {j_2}_{-\mu})+\delta_{- \epsilon}\mu \epsilon {b_2}_{-\mu} }{\sqrt{2}\sqrt{1+\nu\mu {j_2}_{-\mu}}} \left| \beta_{\mu \cdot \epsilon \epsilon} \right> \\ \nonumber
\left| \phi_{\mu \nu}^{3} \right> =&\sum_{\epsilon \in \{-,+\}} \frac {(1+\nu {b_3}_{-\mu})+\nu \epsilon {j_3}_{-\mu} }{2\sqrt{1+\nu {b_3}_{-\mu}}} \left| \beta_{\epsilon \mu} \right> \nonumber \\
\end{align}

\noindent where $\delta_{\alpha \beta}$ is the Kronecker delta. For further applications, note that in this notation, a general bipartite state can be written as in computational basis as in Bell basis as:

\begin{equation}\label{twoqubit}
\left| \psi \right>=\sum_{{A,B} \in \{ 0, 1 \}} \mathcal{A}_{A B} \left| A B \right> = \sum_{{\alpha \beta} \in \{-, + \}} \mathcal{B}_{\alpha,\beta} \left| \beta_{\alpha \beta} \right> 
\end{equation}

\subsection{Evolution operator solutions}

Using analytical expressions for eigenvalues and eigenvectors, we adopt the following reduced definitions associated with the energy levels:

\begin{equation} \label{delta}
{\Delta_h}_\mu^\nu = \frac{t}{2} ({E_h}_{\mu +}+\nu {E_h}_{\mu -})=
\begin{cases}
\mu J_h t & \rm{if} \quad \nu=+ \\
{R_h}_{-\mu} t & \rm{if} \quad \nu=-
\end{cases}
\end{equation}

\noindent and the variables:

\begin{eqnarray} \label{ed}
{e_h}_\alpha^\beta &=& \cos {\Delta_h}_\alpha^- + i \beta {j_h}_{-\alpha} \sin {\Delta_h}_\alpha^- \\ \nonumber
{d_h}_\alpha &=& {b_h}_{-\alpha} \sin {\Delta_h}_\alpha^-
\end{eqnarray}

\noindent then the evolution operator in Bell basis:

\begin{equation} \label{evop1}
U_{h}(t)=\sum_{\alpha,\beta,\gamma, \delta} {U_h}_{\alpha\beta,\gamma\delta} \left| \beta_{\alpha \beta} \right>\left< \beta_{\gamma \delta} \right|
\end{equation}

can be written in matrix form as:

\begin{widetext}
\begin{eqnarray} \label{mathamiltonian}
{U_{1}}(t)=& \left(
\begin{array}{c|c|c|c}
{e^{i {\Delta_1}_-^+}{e_1}_-^-}^* & i e^{i {\Delta_1}_-^+}{d_1}_-      & 0           & 0      \\
\hline
i e^{i {\Delta_1}_-^+}{d_1}_- & e^{i {\Delta_1}_-^+}{{e_1}_-^-}  & 0           & 0            \\
\hline
0         & 0          & {e^{i {\Delta_1}_+^+}{e_1}_+^+}^*  & -i e^{i {\Delta_1}_+^+}{d_1}_+  \\
\hline
0         & 0          & -i e^{i {\Delta_1}_+^+}{d_1}_+     & {e^{i {\Delta_1}_+^+}{e_1}_+^+} 
\end{array}
\right)  &\in \mathbb{S}_1 \\[3mm] \nonumber
{U_{2}}(t)=& \left(
\begin{array}{c|c|c|c}
e^{i {\Delta_2}_+^+}{{e_2}_+^+}^*    &   0   &   0   & - e^{i {\Delta_2}_+^+}{d_2}_+  \\
\hline
0  &  e^{i {\Delta_2}_-^+}{{e_2}_-^+}^* &  -e^{i {\Delta_2}_-^+}{{d_2}_-}  & 0        \\
\hline
0  &  e^{i {\Delta_2}_-^+}{{d_2}_-} &  e^{i {\Delta_2}_-^+}{{e_2}_-^+}  & 0           \\
\hline
e^{i {\Delta_2}_+^+}{d_2}_+    &   0   &   0   & e^{i {\Delta_2}_+^+}{{e_2}_+^+}  
\end{array} 
\right) &\in \mathbb{S}_2 \\[3mm] \nonumber
{U_{3}}(t)=& \left(
\begin{array}{c|c|c|c}
e^{i {\Delta_3}_-^+}{{e_3}_-^+}^* & 0 & i e^{i {\Delta_3}_-^+}{d_3}_-      & 0        \\
\hline
0  &  e^{i {\Delta_3}_+^+}{{e_3}_+^+}^* & 0 & i e^{i {\Delta_3}_+^+}{d_3}_+           \\
\hline
i e^{i {\Delta_3}_-^+}{d_3}_- & 0 &  e^{i {\Delta_3}_-^+}{{e_3}_-^+}      & 0         \\
\hline
0  &  i e^{i {\Delta_3}_+^+}{d_3}_+ & 0 & e^{i {\Delta_3}_+^+}{{e_3}_+^+} 
\end{array}
\right) &\in \mathbb{S}_3
\end{eqnarray}
\end{widetext}

As is clarified in \cite{delgadoA} to avoid misconceptions, note that scripts $-,+$ used in last definitions and results are related with energy labels more than internal operations (as those previously defined, ${J_{\{h\}}}_\pm,{B_{\{h\}}}_\mp$). 

$U_{h}(t)$ clearly exhibit a $2 \times 2$ sector structure which is seminal in current work. As was stated in \cite{delgadoA}, this sector is an element of $U(2)=U(1) \times SU(2)$ with reciprocal determinants between sectors to conform the $4 \times 4$ structure of evolution operator in $SU(4)$. As was stated in \cite{delgadoA}, $\mathbb{S}_1, \mathbb{S}_2, \mathbb{S}_3$:

\begin{eqnarray}\label{subgroups}
\mathbb{S}_1 &=& \{ A \in SU(4) | A_{\alpha \beta, \gamma \delta}= \delta_{\alpha \gamma} u_{\alpha \beta, \gamma \delta}, \nonumber \\  
&& \quad (u_{\gamma \alpha, \gamma \beta})\vline_{\gamma=\pm} \in U(2)\} \\ \nonumber
\mathbb{S}_2 &=& \{ A \in SU(4) | A_{\alpha \beta, \gamma \delta}= \delta_{\alpha \cdot \gamma \beta \cdot \delta} u_{\alpha \beta, \gamma \delta} , \nonumber \\  
&& \quad (u_{\alpha \beta, \gamma \cdot \alpha \gamma \cdot \beta})\vline_{\gamma=\pm} \in U(2)\} \\ \nonumber
\mathbb{S}_3 &=& \{ A \in SU(4) | A_{\alpha \beta, \gamma \delta}= \delta_{\beta \delta} u_{\alpha \gamma, \beta \gamma} , \nonumber \\  
&& \quad (u_{\gamma \alpha, \gamma \beta})\vline_{\gamma=\pm} \in U(2)\} \\ \nonumber
\end{eqnarray}

\noindent are subgroups of $SU(4)$. In addition, we state ${\mathbb S}^*_h \subset {\mathbb S}_h$ to each set of matrices able to be generated by $U_h(t)$ in (\ref{mathamiltonian}). As a result of that work, this subset for each specific physical parameters ${j_h}_\pm, {b_h}_\pm$ is again a subgroup of ${\mathbb S}_h$. Thus, inverses of last operator, $U^\dagger_h(t)$, can be obtained as another $U_h(t')$ for same physical parameters ${j_h}_\pm, {b_h}_\pm$ in the system). This aspect is important for evolution loops because it implies that it can be achieved in at least two pulses of Hamiltonian (\ref{hamiltonian}).
\subsection{Sectors structure}

An important aspect in current work will be state the general structure for  each $2 \times 2$ sectors in $U(2)$:

\begin{equation}\label{sector}
{s_h}_{j} = {e^{i {\Delta_h}_\alpha^+} \left(
\begin{array}{cc}
{{e_h}_\alpha^\beta}^* & -q i^h {d_h}_\alpha   \\
q {i^*}^h {d_h}_\alpha & {{e_h}_\alpha^\beta}    
\end{array}
\right) \vline}_{\tiny \begin{aligned} \alpha &= (-1)^{h+j+1} \\ \beta &= (-1)^{j(h+l_j-k_j+1)} \\ q &= \beta (-1)^{h+1} \end{aligned}}
\end{equation}

\noindent where $h$ is the associated spatial coordinate of magnetic field, $j=1, 2$ is an ordering label for sector as it appears in the rows of the evolution matrix, corresponding with $k_j, l_j$, the labels for its rows in each matrix of (\ref{mathamiltonian}) (by example, $k_2=3, l_2=4$ are the labels for the second sector, $j=2$, in $U_{h=1}(t)$, it means ${s_2}_1)$. 

Note that $\det ({s_h}_{j})={e^{2i {\Delta_h}_\alpha^+}}$ and that nevertheless ${s_h}_j \in U(2)$, not all element of $U(2)$ is a ${s_h}_j$ sector. In fact, ${s_h}_j$ is not necessarily a subgroup of $U(2)$, so if two or more sectors with different physical parameters ${j_h}_\pm, {b_h}_\pm$ are combined in a product, it has not closure, which open opportunities to extend their coverage in $U(2)$ with two o more pulses.

\begin{figure}[th]
\begin{picture}(440,230)(-10,-5)
\put(0,0){\makebox(220,115){\vspace*{4cm}
\scalebox{0.68}[0.68]{
\includegraphics{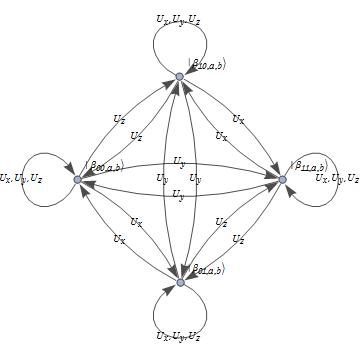}}}}
\end{picture}
\caption{\label{fig1} Graph containing some control transformations between Bell states for qubits $a$ and $b$ which can be obtained by different control operations reducible to forms (\ref{evloop}) and  (\ref{exch}) through magnetic pulses in specific directions.}
\end{figure}

As was suggested in \cite{delgadoA}, by combining several adequate interactions it is possible get the following basic combined forms for sectors in $t=T$ (named diagonalization and antidiagonalization cases respectively):

\begin{eqnarray} 
{s_h}_{j} &=& \pm \mathbb{I}_2 \label{evloop} \\ \nonumber \\
{s_h}_{j}=\pm {\sigma_1} \quad &{\rm or:}& \quad {s_h}_{j}=\pm i \sigma_2 \label{exch}
\end{eqnarray}

\noindent where we need avoid a confusion between operators in computational basis and Bell basis about the use of Pauli matrices $\sigma_1,\sigma_2$ in last expressions, which are stated only as desired forms in the matrix sector. Then we can achieve evolution loops \cite{mielnik1, fernandez1, delgado3, delgado4} and exchange operations \cite{delgado2} in $\mathcal{H}^{\otimes 2}$, switching or recovering any Bell state into another, which will lets obtain several main control effects because it is possible manipulate Bell states in a programmed way by applying a sequence of magnetic field pulses in different directions as is shown in Figure \ref{fig1}, setting transitions or loops between them.

\section{Representation space of evolution operations}
\subsection{Two qubit space and restricted space of representations}\label{section4a}

Space in which a two qubit system (\ref{twoqubit}) evolves is the sphere of seven dimensions $S^7$ (or $S^6$ if we neglect the global phase), with $\mathcal{A}_{AB} \in \mathbb{C}$ and $A,B \in \{ 0,1 \}$. Because high dimensionality of this space to show graphically the evolution of operations depicted, it is more convenient do it in a reduced space. For this purpose, we define:

\begin{eqnarray} \label{requbit}
\left| \psi \right> &=& \sin \alpha \sin \beta \cos \gamma \left| 00 \right> + \sin \alpha \sin \beta \sin \gamma \left| 01 \right> + \nonumber \\
& & \quad \sin \alpha \cos \beta \left| 11 \right> + \cos \alpha \left| 10 \right>\nonumber
\\  \\
{\rm with:} & & \alpha, \beta, \gamma \in \left[0, 2 \pi \right) \nonumber
\end{eqnarray}

\noindent which, as a general quantum state for a two level bipartite system, generate a subspace in ${\mathcal H}^{\otimes 2}$ that we call representation space in the following. This space has some specific properties. If we compare states: $(\alpha=\pi+\epsilon_\alpha, \beta, \gamma), 0<\epsilon_\alpha, 0\le\beta\le\pi, 0\le\gamma\le\pi$ and $(\alpha'=\pi-\epsilon_\alpha, \beta'=\pi-\beta, \gamma'=\gamma+\pi)$ are the same quantum state. Similarly, states: $(\alpha, \beta=\pi+\epsilon_\beta, \gamma), 0\le\alpha\le\pi,0<\epsilon_\beta, 0\le\gamma\le\pi$ and $(\alpha, \beta'=\pi-\epsilon_\beta, \gamma'=\gamma+\pi)$ are equivalent. It shows that representation state can be restricted to $\mathcal{R}=(\alpha,\beta,\gamma) \in [0,\pi] \times [0, \pi] \times [0,2\pi)$ so, $\mathcal{R}$ is a chart on $S^3 \subset S^7$, the three dimensional sphere embed in four dimensions. This space will be used to illustrate the control operations (\ref{evloop}) and (\ref{exch}).

Furthermore, we can note that transformation: $(\alpha, \beta, \gamma) \rightarrow (\alpha', \beta', \gamma')$, with: $0 \le \alpha <\pi, 0 \le \beta<\pi, 0 \le \gamma<\pi$ and $(\alpha'=\pi-\alpha, \beta'=\pi-\beta, \gamma'=\gamma+\pi)$, changes $\left| \psi \right>$ into $-\left| \psi \right>$, which represent same states. With this, representation space can be limited to the cube $\Hat{\mathcal{R}}=\alpha,\beta,\gamma \in [0,\pi)^{\times 3}$. This last structure will be used in the following at least that graphical interpretation will be unclear. Last property in $\Hat{\mathcal{R}}$ provides it a M\"{o}bius like topology in this space, related with its edges identification, an structure inherited from arbitrariness of quantum state phase (the same structure for single qubit states, which is reduced by using $\frac{\theta}{2}$ in their component expressions on Bloch sphere, polar angle is used as $\frac{\theta}{2}$ in component expressions).

\subsection{Entanglement properties of $\Hat{\mathcal{R}}$ and $S^7$ projections on $S^3$}
By calculating the Schmidt coefficients $\lambda_{\pm}$ of (\ref{requbit}), we can obtain the concurrence $\mathcal{C}$:

\begin{eqnarray} \label{concurrence}
\mathcal{C}^2&=&4 \sin^2\alpha \sin^2\beta (1- \sin^2\alpha \sin^2\beta-  \nonumber \\
& & \quad (\cos \alpha \cos \gamma +\sin \alpha \sin \gamma \cos \beta)^2) 
\end{eqnarray}

Thus, $\mathcal{C}=1$ define the maximal entanglement states and $\mathcal{C}=0$ defines separable states. Clearly the lateral edges of $\Hat{\mathcal{R}}$, with $\alpha, \beta=0,\pi$ are separable states, but they are not exclusive, another inner surface in $\Hat{\mathcal{R}}$ contains additional separable states. 

States (\ref{twoqubit}) in $S^7$ could be projected on $S^3$ in several ways. By example if we take:

\begin{eqnarray} \label{proj1}
\alpha&=&\arccos |\mathcal{A}_{1 0}| \nonumber \\
\beta&=&\arccos \frac{|\mathcal{A}_{1 1}|}{\sin \alpha} \nonumber \\
\gamma&=&\arccos \frac{|\mathcal{A}_{0 0}|}{\sin \alpha \sin \beta} 
\end{eqnarray}

\noindent because $\left< \psi | \psi \right>=\sum_{{A,B} \in \{ 0, 1 \}} |\mathcal{A}_{A B}|^2=1$, it lets understand $S^7=S^3 \times (S^1)^{\times 4}$, where each $S^1$ is the fiber bundle corresponding with phases of $\mathcal{A}_{A B}$. This sphere is constructed with intersection points between $S^7$ and subspace constructed with direction of Argand representation for each state component in computational basis on Fock space $\mathcal{H}^{\otimes 2}$ (for instantaneous state). Of course, this projection maps states only on one eighth of $\Hat{\mathcal{R}}$.  Nevertheless, in order to realize better the evolution trajectories in $\Hat{\mathcal{R}}$, we will take a projection based on real parts of components instead of magnitudes (a similar structure was recently studied in \cite{wie1} based on a quaternion construction):

\begin{eqnarray} \label{proj2}
\alpha&=&\arccos  \rm{Re}\mathcal{A}_{1 0} \nonumber \\
\beta&=&\arccos \frac{\rm{Re}\mathcal{A}_{1 1}}{\sin \alpha} \nonumber \\
\gamma&=&\arccos \frac{\rm{Re}\mathcal{A}_{0 0}}{\sin \alpha \sin \beta}
\end{eqnarray}

\noindent this selection sets a maximal $S^3$ sphere on $S^7$ defined by intersection points of real axes of each component in computational basis of $\mathcal{H}^{\otimes 2}$ with $S^7$. 

\begin{figure}[th]
\begin{picture}(440,230)(-10,20)
\put(0,0){\makebox(220,115){\vspace*{5cm}
\scalebox{0.67}[0.67]{
\includegraphics{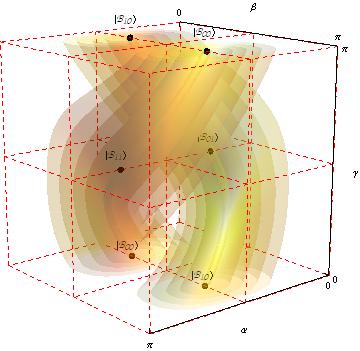}}}}
\end{picture}
\caption{\label{fig2} Representation space $\Hat{\mathcal{R}}$ for two qubits depicted in text. Note where Bell states are located and how space is twisted relating points in low and upper edges, repeating itself vertically and horizontally. Thus, vertical edges, $\alpha, \beta=0,\pi$, extend space in lateral directions as reflexions respect themselves. Level surfaces for concurrence are shown, inner ones approaching to the maximal entanglement line, $\mathcal{C}=1$, where Bell states lie. Vertical external faces in $\Hat{\mathcal{R}}$ contains separable states where $\mathcal{C}=0$, together with an inner surface splitting the two tube-like coverings.}
\end{figure}

\section{Control operations}

\subsection{Generalities}

As is well known, one of the goals in control of Ising model is obtain control on Rabi oscillations generated in this case for inhomogeneous and anisotropic field and interaction strengths respectively. Normally, this manipulation involves a tight control of time \cite{kiri1,meng1} and alternative types of control, as resonant or on-off control \cite{cruz1, cruz2, medina1}.

In this section we will analyze how evolution operators sectors could adopt specific forms (\ref{evloop}) and (\ref{exch}) in order to become into practical control on Bell states through evolution loops and exchange operations. These control effects based on Bell states could be involved in applications as quantum characterization, teleportation, discrimination and repreparation. Because (\ref{hamiltonian}) generalizes particular interactions included in other works, some of the following results could be applied to generalize those situations. Note that this kind of control operations could be applied in several ways, not only as control scheme but as correction scheme. As instance, operation (\ref{evloop}), could be useful for some quantum state $\left| \psi \right>$ (neither entangled or separable), when it has been distorted by a magnetic field. Then, knowledge about evolution loops is useful to restore it under same field conditions \cite{delgado1, delgado2}:

\begin{eqnarray} \label{restoreloop}
U_h(t_2)U_h(t_1)\left| \psi \right>=\left| \psi \right>
\end{eqnarray}

\noindent by measuring accurately distortion time $t_1$ and restoring time $t_2$. Similarly,  operation (\ref{exch}) can be used not only to transform Bell states between them, but to reprepare a distorted Bell state into other:

\begin{eqnarray} \label{restoreexch}
U_h(t_2)U_h(t_1)\left| \beta_{\epsilon \delta} \right>=\left| \beta_{\epsilon' \delta'} \right>
\end{eqnarray} 

\subsection{Sector conditions for diagonalization and antidiagonalization}

\subsubsection{Diagonalization and antidiagonalization: one pulse solutions}

In order to ${s_h}_{j}$ fulfills (\ref{evloop}) and (\ref{exch}), there are several possibilities. The more basic way is to use one field pulse. An ease and direct analysis shows that ${s_h}_{j}=\pm \mathbb{I}_2$ in only one pulse if the following prescriptions are fulfilled:

\begin{eqnarray} \label{evloop1}
T&=&\frac{m_\alpha-n_\alpha}{\alpha J_h}\pi >0 \\ \nonumber
{B_h}_{-\alpha}^2&=&(\frac{J_h n_\alpha}{m_\alpha-n_\alpha})^2-{{J_{\{h\}}}_\alpha}^2 \\ \nonumber
\\ \nonumber
{\rm with:} && n_\alpha,m_\alpha \in \mathbb{Z} \\ \nonumber
\end{eqnarray}

Still, prescriptions (\ref{evloop1}) should be combined and reviewed if they are compatible in both sectors of each model (\ref{mathamiltonian}) to obtain an evolution loop $U_h=(-1)^{m_\alpha}\mathbb{I}_4$. It is possible if $m_\alpha-n_\alpha=n_{-\alpha}-m_{-\alpha}$.

If condition (\ref{exch}) is included in at least one of sectors then $U_h$ becomes an exchange operation which lets transform some Bell states into other. Nevertheless, there are not feasible solutions in one pulse for sector antidiagonalization (${s_h}_{j}=\pm \sigma_2$) without to restrict $J_h$ and to use a finite external field, then a multiple pulse approach is necessary. As we will see, there are solutions in two pulses to obtain diagonal-antidiagonal evolution matrices. 

\subsubsection{Diagonalization and antidiagonalization: two pulses solutions}

Because sector properties of evolution matrices (\ref{mathamiltonian}), considering two pulses with external field in same direction
requires to fit with each form (\ref{evloop}) and (\ref{exch}) the matrix sector obtained by multiplying two basic sectors corresponding to these two consecutive pulses, ${s'_h}_j$ after of ${s_h}_j$ (clearly $\alpha, \beta, q$ are the same for both pulses in the same sector):

\begin{widetext}
\begin{eqnarray} \label{twopulses}
{s'_h}_{j} {s_h}_{j}={e^{i ({\Delta'_h}_\alpha^+ + {\Delta_h}_\alpha^+)}} \times  
 \left(
\begin{array}{cc}
{{e'_h}_\alpha^\beta}^*{{e_h}_\alpha^\beta}^* - {d'_h}_\alpha {d_h}_\alpha & 
-q i^h ({{e'_h}_\alpha^\beta}^*{d_h}_\alpha + {{e_h}_\alpha^\beta}{d'_h}_\alpha )   \\
q {i^*}^h ({{e'_h}_\alpha^\beta}{d_h}_\alpha + {{e_h}_\alpha^\beta}^*{d'_h}_\alpha ) & 
{{e'_h}_\alpha^\beta}{{e_h}_\alpha^\beta} - {d'_h}_\alpha {d_h}_\alpha    
\end{array}
\right)
\end{eqnarray}
\end{widetext}

In order to diagonalize or antidiagonalize last two pulse generic sector, some general conditions should be fulfilled. In terms of definitions (\ref{defs}), (\ref{ed}) and noting that $|{e_h}_\alpha^\beta|^2+|{d_h}_\alpha|^2=1$, we obtain for diagonal form:

\begin{eqnarray}\label{diag}
& {j_h}_{-\alpha}\tan {\Delta_h}_\alpha^{-}+{j'_h}_{-\alpha} \tan {\Delta'_h}_\alpha^{-} = 0 \\ \nonumber
& {\rm sign} ({b_h}_{-\alpha}{b'_h}_{-\alpha}\sin {\Delta_h}_\alpha^- \sin{\Delta'_h}_\alpha^- \cos {\Delta_h}_\alpha^- \cos {\Delta'_h}_\alpha^-) = -1 \\ \nonumber
&|{e_h}_\alpha^\beta| |{d'_h}_\alpha|=|{e'_h}_\alpha^\beta| |{d_h}_\alpha| \rightarrow |{e_h}_\alpha^\beta| = |{e'_h}_\alpha^\beta|, |{d'_h}_\alpha| = |{d_h}_\alpha|
\end{eqnarray}

\noindent and similarly for antidiagonal one:

\begin{eqnarray}\label{adiag}
& {j_h}_{-\alpha}\tan {\Delta_h}_\alpha^{-}+{j'_h}_{-\alpha} \tan {\Delta'_h}_\alpha^{-} = 0 \\ \nonumber
& {\rm sign} ({b_h}_{-\alpha}{b'_h}_{-\alpha} \sin {\Delta_h}_\alpha^- \sin{\Delta'_h}_\alpha^- \cos {\Delta_h}_\alpha^- \cos {\Delta'_h}_\alpha^-) = 1 \\ \nonumber
&|{e_h}_\alpha^\beta| |{e'_h}_\alpha^\beta| = |{d_h}_\alpha||{d'_h}_\alpha| \rightarrow |{e_h}_\alpha^\beta| = |{d'_h}_\alpha|, |{e'_h}_\alpha^\beta|  = |{d_h}_\alpha|
\end{eqnarray}

Note in (\ref{diag}) and (\ref{adiag}) that first requirement is the same in both cases; second equation in each set just adjust signs for the necessary combination between ${b_h}_{-\alpha}, {b'_h}_{-\alpha}$; finally, third expressions are equivalent to:

\begin{eqnarray}
{b_h}_{-\alpha}^2 \sin^2 {\Delta_h}_\alpha^-={b'_h}_{-\alpha}^2 \sin^2 {\Delta'_h}_\alpha^-
\end{eqnarray}

\noindent for diagonalization and:

\begin{eqnarray}
{b_h}_{-\alpha}^2 \sin^2 {\Delta_h}_\alpha^- +{b'_h}_{-\alpha}^2 \sin^2 {\Delta'_h}_\alpha^-=1 
\end{eqnarray}

\noindent for antidiagonalization. Combining these expressions with the first equations in (\ref{diag}) and (\ref{adiag}), we obtain different solutions. With these conditions fulfilled, sector becomes for diagonalization and antidiagonalization respectively:

\begin{eqnarray} \label{interdiagadiag}
 {e^{i ({\Delta'_h}_\alpha^+ + {\Delta_h}_\alpha^+)}} {\rm sign} (\cos {\Delta_h}_\alpha^- \cos{\Delta'_h}_\alpha^-) \mathbb{I}
\\ \nonumber  \\ 
 q {i}^h {e^{i ({\Delta'_h}_\alpha^+ + {\Delta_h}_\alpha^+)}} {\rm sign} (\cos {\Delta_h}_\alpha^- \sin{\Delta'_h}_\alpha^- {b'_h}_{-\alpha}) 
\times \\ 
 \nonumber \left(
\begin{array}{cc}
0 & 
- e^{i \beta {\varphi_h}_\alpha}  \\
(-1)^h e^{i \beta {\varphi'_h}_{-\alpha}} & 
0    
\end{array}
\right) 
\end{eqnarray}

\noindent with:

\begin{eqnarray}
{\varphi_h}_\alpha &=& \arctan ({j_h}_{-\alpha} \tan {\Delta_h}_\alpha^{-}) \\ \nonumber 
{\varphi'_h}_\alpha &=& \arctan ({j'_h}_{-\alpha} \tan {\Delta'_h}_\alpha^{-})
\end{eqnarray}

The more feasible solution for diagonalization case is:

\begin{eqnarray}\label{diag2}
&{\Delta_h}_\alpha^- +{\rm sign}({J_{\{h\}}}_{\alpha}{J'_{\{h\}}}_{\alpha}) {\Delta'_h}_\alpha^-=n_\alpha \pi \\ \nonumber
&{\Delta_h}_\alpha^+ + {\Delta'_h}_\alpha^+= (m_\alpha+n_\alpha) \pi \\ \nonumber
&\frac{{B_h}_{-\alpha}}{{J_{\{h\}}}_\alpha}=\frac{{B'_h}_{-\alpha}}{{J'_{\{h\}}}_\alpha} \\ \nonumber
{\rm with:} & m_\alpha, n_\alpha \in \mathbb{Z}
\end{eqnarray}

\noindent giving exactly:

\begin{eqnarray} \label{identity}
{s'_h}_{j} {s_h}_{j}= 
 (-1)^{m_\alpha}\mathbb{I}_2
\end{eqnarray}

Similarly, for antidiagonalization the more feasible solution is (obtained as limit case of (\ref{adiag}) or directly from (\ref{twopulses})):

\begin{eqnarray}\label{adiag2}
&{\Delta_h}_\alpha^- =\frac{2n_\alpha+1}{2} \pi, {\Delta'_h}_\alpha^-=\frac{2n'_\alpha+1}{2} \pi \\ \nonumber
&{\Delta_h}_\alpha^+ + {\Delta'_h}_\alpha^+ = - \frac{\pi}{2} (h+{\rm sign} (q \beta b'_{h_{-\alpha}}j_{h_{-\alpha}}) + \quad \\ \nonumber 
& \quad \quad \quad 2(n_\alpha+n'_\alpha-s_\alpha+1))  \\ \nonumber
& \quad \equiv \frac{\pi}{2} M_{h, q,\alpha, \beta, n_\alpha, n'_\alpha,s_\alpha} \quad \\ \nonumber
&\frac{{B_h}_{-\alpha}}{{J_{\{h\}}}_\alpha}=-\frac{{J'_{\{h\}}}_\alpha}{{B'_h}_{-\alpha}} \\ \nonumber
{\rm with:} & s_\alpha, n_\alpha, n'_\alpha \in \mathbb{Z}  
\end{eqnarray}

\noindent giving: 

\begin{eqnarray} \label{adiag2p}
{s'_h}_{j} {s_h}_{j} &=& 
\left(
\begin{array}{cc}
0 & 
(-1)^{s_\alpha}  \\
(-1)^{h+s_\alpha} & 
0    
\end{array}
\right) \nonumber \\
&=& (-1)^{s_{\alpha}}i^{h  {\rm mod}  2} \sigma_{1+h  {\rm mod}  2}
\end{eqnarray} 

\noindent noting that evolution matrix ${s'_h}_{j} {s_h}_{j}$ reduces to (\ref{exch}). Signs induced by $s$ and $h$ introduce a phase when associated Bell states are exchanged. Note that (\ref{diag2}) and (\ref{adiag2p}) should not be combined for same sector, instead each condition is used on one different sector, one with $\alpha=-1$ and another with $\alpha=1$ at time, in agreement with (\ref{sector}). Anyway, each set of conditions fitting only one sector of whole evolution matrix, so compatibility between them should be adjusted.

\subsection{Evolution loops and exchange operations}

\subsubsection{Evolution loops in one pulse}

As was mentioned before, evolution loops can be reached with only one field pulse with the prescriptions given in (\ref{evloop1}) combined for two sectors (labeled as $\pm \alpha$):

\begin{eqnarray} \label{evloop2}
T&=&\frac{m_\alpha-n_\alpha}{\alpha J_h}\pi=\frac{n_{-\alpha}-m_{-\alpha}}{\alpha J_h}\pi >0 \\ \nonumber
{B_h}_{-\alpha}^2&=&(\frac{J_h n_\alpha}{m_\alpha-n_\alpha})^2-{{J_{\{h\}}}_\alpha}^2 >0 \\ \nonumber
{B_h}_{\alpha}^2&=&(\frac{J_h n_{-\alpha}}{m_{-\alpha}-n_{-\alpha}})^2-{{J_{\{h\}}}_{-\alpha}}^2 >0 \\ \nonumber
\\ \nonumber
{\rm with:} && n_{\pm \alpha}, m_{\pm \alpha} \in \mathbb{Z} 
\end{eqnarray}

\noindent where $n_{\pm \alpha}, m_{\pm \alpha}$ should be properly selected. If additionally, $m_{\pm \alpha}$ have the same parity, then $U_h(t)$ will have the same phase in both sectors. Thus, these prescriptions generate the matrix evolution:

\begin{equation}
U_{h}(t)=(-1)^{m_\alpha} \mathbb{I}_4
\end{equation}

\begin{figure}[th]
\begin{picture}(440,240)(-10,35)
\put(0,0){\makebox(220,120){\vspace*{6cm}
\scalebox{0.56}[0.56]{
\includegraphics{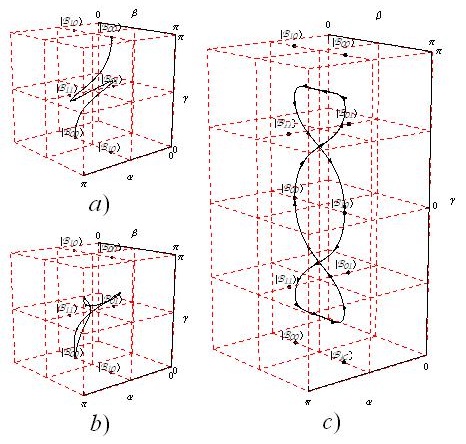}}}}
\end{picture}
\caption{\label{fig3} Representation of evolution loops in $\Hat{\mathcal{R}}$ for Bell state $\left| \beta_{00} \right>$ for each type of Hamiltonian (\ref{hamiltonian}): a) $U_x$ with $m_{-}=2,n_{-}=1,m_{+}=1,n_{+}=2$; b) $U_y$ with $m_{-}=4,n_{-}=1,m_{+}=2,n_{+}=5$; and $U_z$ with $m_{-}=4,n_{-}=1,m_{+}=2,n_{+}=5$. All trajectories begin and end in $\left| \beta_{00} \right>$.}
\end{figure}

Figure \ref{fig3} shows three cases of evolution loops represented on $\Hat{\mathcal{R}}$ for $\left| \beta_{00} \right>$ for $J_i = 10, J_j = 0.4, J_k = 0.5$, where $i, j, k$ is an even permutation of $1, 2, 3$, and $i$ corresponding with $1, 2, 3$ or $x, y, z$ respectively, is the direction of magnetic field applied in each example. Each figure a, b and c, shows the evolution loop departing from $\left| \beta_{00} \right>$ and generated by $U_x, U_y, U_z$ respectively. In that representation, reader should remember that evolution trajectory is only a projection from whole space onto $S^3$, so trajectory does not necessarily show the complete real states (specially when it crosses other Bell states, as in Figure \ref{fig3}c), except in its edges, which corresponds with $\left| \beta_{00} \right>$ state. In these representations, continuity was preferred, so trajectory in Figure \ref{fig3}c was extended outside $\Hat{\mathcal{R}}$ (otherwise, lower part of trajectory will appear from top of $\Hat{\mathcal{R}}$, but inverted with respect to $\beta$ direction to finally arrive on $\left| \beta_{00} \right>$, in agreement with symmetries stated in section \ref{section4a}).

\subsubsection{Evolution loops in two pulses}

For this case, we identify six free possible parameters: ${B_h}_{\pm \alpha}, {B'_h}_{\pm \alpha}, t, t'$. Nevertheless is possible to achieve evolution loops (\ref{evloop}) with two pulses, these operations are complex and unnecessary in spite of last results. In addition, some of them are possible as combinations of one pulse operations (by example, $J_{\pm \alpha}=J'_{\pm \alpha}$ case reduces exactly to one pulse operations), so we will omit this analysis. 

There are some possible useful issues around reversibility which is convenient remark here. Note that if we split arbitrarily the process of last subsection in two times $t+t' = T$, we obtain exactly two inverse operations between them (we restrict our discussion to $m_\alpha$ even), $U_h(t')U_h(t) = \mathbb{I}_4 \rightarrow U^{-1}_h(t) = U_h(t')$. This inverse operation fulfills for both sectors,  but one can be interested in to reverse selectively the evolution in only one sector and pursuit different effects in the remaining sector. It denotes that evolution operations can involve their own inverse operations at least in special and controlled cases. In the next subsection, we will focus on two pulses operations to construct exchange operation solutions: diagonal-antidiagonal or antidiagonal-antidiagonal sectors.

\subsubsection{Exchange operations in two pulses}

It is easy see that general sector (\ref{sector}) can be antidiagonalized while magnetic field remains finite (the only way that $|{e_h}^\beta_\alpha|=0$), thus, the attempt is use two pulses. Clearly, in spite of (\ref{adiag2}), the case when evolution matrices (\ref{mathamiltonian}) become in a antidiagonal-antidiagonal case for both sectors implies an extra condition on (\ref{diag2}), which requires strong restrictions on interaction strengths $J_i$. For this reason, we will analyze only the diagonal-antidiagonal case, where two selected Bell state could be exchanged in agreement with diagram of Figure \ref{fig1}. Combining last solution, antidiagonal-antidiagonal case can be reached in four pulses. 

Last solution requires combine (\ref{diag2}) and (\ref{adiag2}) equations for two sectors in any case of (\ref{mathamiltonian}). A solution is reached by setting a program based on calculate each one of ${B_h}_{-\alpha}, {B'_h}_{\pm \alpha}, t, t'$ parameters in terms of ${B_h}_{\alpha}$. Here, we set $\alpha$ script for diagonal sector and $-\alpha$ for antidiagonal one. First, it is possible decouple ${B_h}_\alpha$ in the following equation:

\begin{eqnarray} \label{diagadiag1}
\left|\frac{{B_h}_{\alpha}}{{J_{\{h\}}}_{-\alpha}}\right|^2+1&=& \left(A+B\left|\frac{{B_h}_\alpha}{{J_{\{h\}}}_{-\alpha}}\right|\right)^2 \nonumber \\ \\
{\rm with:} && A=\frac{(2n_{-\alpha}+1)J_h}{2(m_\alpha+n_\alpha)|{J_{\{h\}}}_{-\alpha}|} \nonumber \\ 
&& B=\frac{(2n'_{-\alpha}+1)J'_{h}}{2(m_\alpha+n_\alpha)|{J'_{\{h\}}}_{-\alpha}|} \nonumber
\end{eqnarray}

\begin{figure}[th]
\begin{picture}(440,350)(-10,20)
\put(0,0){\makebox(220,175){\vspace*{7cm}
\scalebox{0.67}[0.67]{
\includegraphics{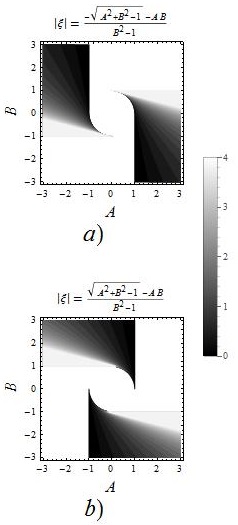}}}}
\end{picture}
\caption{\label{fig4} Solutions for equation (\ref{diagadiag1}) in the plane $A - B$. a) Negative sign, b) Positive sign. Shaded regions correspond to $|\xi|$ values in the colored scale. Note that solutions are centered in the opposites signs for $A$ and $B$, limiting the spectrum of physical cases in spite of (\ref{diagadiag3}).}
\end{figure}

Equation (\ref{diagadiag1}) it is easily solved, giving the following family of potential solutions:

\begin{eqnarray}\label{solxi}
|\xi|= \frac{-AB\pm\sqrt{A^2+B^2-1}}{B^2-1}
\end{eqnarray}

\noindent where $\xi \equiv \frac{{B_h}_{\alpha}}{{J_{\{h\}}}_{-\alpha}}$. This equation has solutions while $A^2+B^2 \ge 1$ and positivity had been warranted, which is possible in general for finite and anisotropic interaction strengths in three directions by selecting properly values for $n_{-\alpha}, n'_{-\alpha}, n_\alpha, m_\alpha$ (Figure \ref{fig4}).

Clearly sign of $n_{-\alpha}, n'_{-\alpha} \ge 0$ and $m_\alpha + n_\alpha, J_h, J'_h$ determines the selection of $A, B$. Figure \ref{fig4} shows the available regions where (\ref{solxi}) has solutions in plane $A - B$, depicting
$|\xi|$ values in them. Note that there are solutions in
four quadrants which lets find solutions for finite $J_h, J'_h$
with an adequate selection of $n_{-\alpha}, n'_{-\alpha}, n_\alpha, m_\alpha$. Be aware about completely white regions where (\ref{solxi}) have not real solutions. 

Similarly, it is possible write ${B_h}_{-\alpha}$ in terms of last parameter:

\begin{eqnarray} \label{diagadiag2}
\sqrt{{B_h}_{-\alpha}^2+{J_{\{h\}}}_\alpha^2}&=&\frac{2n_\alpha \sqrt{{B_h}_\alpha^2+{J_{\{h\}}}_{-\alpha}^2}}{(2n_{-\alpha}+1)+P_\alpha S_\alpha (2n'_{-\alpha}+1)|{B_h}_\alpha|}  \nonumber \\ \nonumber \\ \\
{\rm with:} && P_\alpha={\rm sign} ({J'_{\{h\}}}_\alpha {J_{\{h\}}}_{ \alpha}) \nonumber \\ 
&& S_\alpha=|\frac{{J'_{\{h\}}}_\alpha}{{J_{\{h\}}}_\alpha {J'_{\{h\}}}_{-\alpha}}| \nonumber
\end{eqnarray}

\noindent so, ${B_h}_\alpha, {B_h}_{-\alpha}$ are determined (with several possibilities for their signs). With that, remaining parameters can be obtained:

\begin{eqnarray} \label{diagadiag3}
{B'_h}_\alpha&=&-\frac{{J_{\{h\}}}_{-\alpha} {J'_{\{h\}}}_{-\alpha}}{{B_h}_\alpha} \nonumber \\
{B'_h}_{-\alpha}&=&\frac{{B_h}_{-\alpha} {J'_{\{h\}}}_{\alpha}}{{J_{\{h\}}}_{\alpha}} \nonumber \\
t&=&\frac{(2n_{-\alpha}+1)\pi}{2\sqrt{{B_h}_\alpha^2+{J_{\{h\}}}_{-\alpha}^2}}  \nonumber \\
t'&=&\frac{(2n'_{-\alpha}+1)\pi }{2\sqrt{{B_h}_\alpha^2+{J_{\{h\}}}_{-\alpha}^2} } \frac{|{B_h}_\alpha|}{|{J'_{\{h\}}}_{-\alpha}|} 
\end{eqnarray}

In addition, selection of parameters involved should fulfill:

\begin{eqnarray}\label{diagadiagphase}
2(m_\alpha+n_\alpha)&=&M_{h, q,-\alpha, \beta, n_{-\alpha}, n'_{-\alpha},s_{-\alpha}}= \nonumber \\
&=&-(h+{\rm sign} (q \beta b'_{h_{\alpha}}j_{h_{\alpha}}) +  \nonumber \\
& & \quad 2(n_{-\alpha}+n'_{-\alpha}-s_{-\alpha}+1))
\end{eqnarray}

\noindent then $m_\alpha +n_\alpha$ fixes $s_\alpha$ value. It is clear that our analysis has been preserving the possibility that those strengths could change during each pulse (a few common situation, but not impossible), but it is not a decisive factor. 

\begin{figure}[th]
\begin{picture}(440,290)(-10,35)
\put(0,0){\makebox(220,145){\vspace*{7cm}
\scalebox{0.65}[0.65]{
\includegraphics{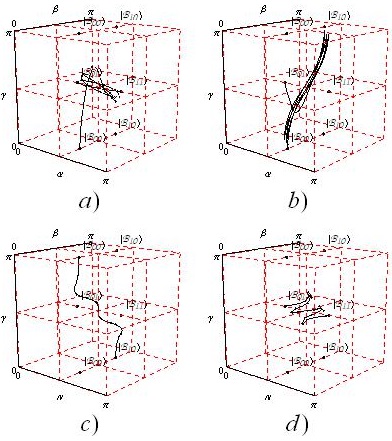}}}}
\end{picture}
\caption{\label{fig5} Evolution obtained as exchange operation for $U_x$
case with first sector as antidiagonal and second as diagonal in (\ref{mathamiltonian}). a) and b) states $\left| \beta_{00} \right>$ and $\left| \beta_{01} \right>$, respectively, interchanging
mutually after a cycle of two pulses constructed in agreement
with prescriptions (\ref{diagadiag1}-\ref{diagadiag3}) under a trajectory in $\Hat{\mathcal{R}}$ developing controlled Rabi oscillations; c) and d) states $\left| \beta_{10} \right>$ and $\left| \beta_{11} \right>$ evolving respectively under same exchange operation, each one on other in a forward-backward trajectory.}
\end{figure}

Figure \ref{fig5} shows the effects of evolution on Bell states 
under Hamiltonian (\ref{hamiltonian}), by applying prescriptions (\ref{diagadiag1}-\ref{diagadiag3})
to generate an exchange operation between Bell states.
In the case shown, with $J_x = 2, J_y = 0.4, J_z = 0.6$
and selecting the first sector as antidiagonal ($j = 1$ in
(\ref{sector})), which implies: $\alpha, \beta, q = -1$. In addition, $n_\alpha =0, n'_\alpha = 0, m_{-\alpha} = 2; n_{-\alpha} = -4$ were selected. These last parameters together with interaction strengths determine Rabi frequencies and magnetic fields involved in the process (until two figures: $t= 1.77, t' = 7.65, {B_1}_- = 1.73, {B_1}_+ = 0.86, {B'_1}_- = 1.73, {B_1}_+= -0.05$), which are reflected on each trajectory in $\Hat{\mathcal{R}}$ through several operations
on Bell states. In agreement with Figure \ref{fig1}, this
selection lets make a transformation between $\left| \beta_{00} \right>$ and $\left| \beta_{01} \right>$ states. Figure \ref{fig5}a and \ref{fig5}b show these trajectories projected in $\Hat{\mathcal{R}}$ from each one of these states directly into
other. While, $\left| \beta_{10} \right>$ and $\left| \beta_{11} \right>$ are preserved under same operation in the same period of time (they do not shown). Figure \ref{fig5}c and \ref{fig5}d show correspondent trajectories projected on $\Hat{\mathcal{R}}$. In these cases, states evolve leaving their original stages, going on far states (which, curiously have projections on $\left| \beta_{00} \right>$ and $\left| \beta_{01} \right>$) and finally they comeback into the original ones. Because $m_{-\alpha} = 2$, states $\left| \beta_{10} \right>$ and $\left| \beta_{11} \right>$ do not exhibit phase change. Instead, because (\ref{diagadiagphase}), $s_\alpha = 10, h = 1$, so only $\left| \beta_{00} \right>$ changes its phase when it evolves into $-\left| \beta_{01} \right>$. Concretely, to fix ideas, for this particular example of these set of operations achievable, evolution matrix becomes in $T = t + t'$, the first matrix of the next set:

\begin{eqnarray} \label{diagadiagforms}
{U_{1}}(T)&=& \left(
\begin{array}{c c c c}
0 & 1 & 0 & 0      \\
-1 & 0 & 0 & 0            \\
0 & 0 & 1 & 0  \\
0 & 0 & 0 & 1 
\end{array}
\right)  , \left(
\begin{array}{c c c c}
1 & 0 & 0 & 0      \\
0 & 1 & 0 & 0            \\
0 & 0 & 0 & 1  \\
0 & 0 & -1 & 0 
\end{array}
\right) \\
{U_{2}}(T)&=& \left(
\begin{array}{c c c c}
0 & 0 & 0 & i      \\
0 & 1 & 0 & 0            \\
0 & 0 & 1 & 0  \\
i & 0 & 0 & 0 
\end{array}
\right) , \left(
\begin{array}{c c c c}
1 & 0 & 0 & 0      \\
0 & 0 & i & 0            \\
0 & i & 0 & 0  \\
0 & 0 & 0 & 1 
\end{array}
\right) \\
{U_{3}}(T)&=& \left(
\begin{array}{c c c c}
1 & 0 & 0 & 0      \\
0 & 0 & 0 & 1            \\
0 & 0 & 1 & 0  \\
0 & -1 & 0 & 0 
\end{array}
\right) , \left(
\begin{array}{c c c c}
0 & 0 & 1 & 0      \\
0 & 1 & 0 & 0            \\
-1 & 0 & 0 & 0  \\
0 & 0 & 0 & 1 
\end{array}
\right)
\end{eqnarray}

\noindent and the last five matrices are other diagonal-antidiagonal forms achievable with this procedure and prescriptions (\ref{diagadiag1}-\ref{diagadiagphase}). Note that similar formulas (with variations in sectors and signs) appear for other parameters, in particular for $h = 1, 3$. For $h = 2$ it is possible select imaginary entries in the diagonal sectors, with which, $U_2(T)$ will have a global imaginary phase (which could be more useful in order to eliminate it). Anyway, as (\ref{adiag2}) implies a phase factor $i$ for $h = 2$ in the antidiagonal sector at least, then the projection (\ref{proj2}) fails to reflect the exchange effects in our operations (or in the evolution loops representation if all entries are taken imaginary), so projections like (\ref{proj1}) are more feasible representations on $\Hat{\mathcal{R}}$. General forms generated by these operations will be discussed in the following subsection as an extended proposal for them.

\subsection{Forms and complementary operations for diagonal and antidiagonal sectors}

Nevertheless that in this work we are interested mainly in evolution loops and exchange operations, at this point is convenient set some extensions of possible forms that combined sectors could achieve. In general terms, evolution matrices (\ref{mathamiltonian}) have possibility to adopt the form of a $P-$unitary matrices, which recently have been reported to achieve factorization of quantum gates \cite{li1}. Nevertheless that this topic will not be discussed here, yet, it is convenient state some operations obtained with more general forms for diagonal and diagonal-antidiagonal matrices as was discussed before (we use here the superscripts $D$ and $A$ to label each case).

\subsubsection{Diagonal-Diagonal forms}

When each set of equations (\ref{diag}) and (\ref{adiag}) are solved, all equations in each set are necessary conditions to make antidiagonal or diagonal entries equal to zero, respectively. For one pulse diagonal case, weakening unitary global phase for $U_h$, we can obtain at time $T$, diagonal matrices with the next form in each diagonal sector ($D_1$):

\begin{eqnarray} \label{graldiag1}
{s_h}_j^{D_1} &=& e^{i \alpha J_h T} (-1)^{n_\alpha} \mathbb{I}_2 \nonumber \\
&\equiv& e^{i \alpha J_h T} {\mathcal S}_{h,j}^{D_1} {\mathbb I}_2
\end{eqnarray}

\noindent if we have the following restriction in each magnetic field (for $\alpha=\pm 1$ corresponding with each sector):

\begin{eqnarray} 
{B_h}_{-\alpha}^2 &=& (\frac{n_\alpha \pi}{T})^2-{J_{\{h\}}}_\alpha ^2
\end{eqnarray}

\noindent where $n_\alpha \in {\mathbb Z}$ and $T$ is a free parameter. This operation is not precisely an evolution loop because it does not assign a global phase to all states, instead, it assigns different phases to Bell states by pairs and then introduces phases in each term of a linear superpositions in other general states.

These procedures let to obtain more general diagonal
forms for $U_h$ than those obtained before, with the structure:

\begin{eqnarray} \label{graldiag2}
{U_{1}}(T)&\equiv& {\mathcal D}_1^\phi = \left(
\begin{array}{c c c c}
S_{1 1}e^{i \phi} & 0 & 0 & 0      \\
0 & S_{1 1}e^{i \phi} & 0 & 0            \\
0 & 0 & S_{1 2}e^{-i \phi} & 0  \\
0 & 0 & 0 & S_{1 2}e^{-i \phi} 
\end{array}
\right) \nonumber \\
{U_{2}}(T)&\equiv& {\mathcal D}_2^\phi = \left(
\begin{array}{c c c c}
S_{2 1}e^{i \phi} & 0 & 0 & 0      \\
0 & S_{2 2}e^{-i \phi} & 0 & 0            \\
0 & 0 & S_{2 2}e^{-i \phi} & 0  \\
0 & 0 & 0 & S_{2 1}e^{i \phi} 
\end{array}
\right) \nonumber \\
{U_{3}}(T)&\equiv& {\mathcal D}_3^\phi = \left(
\begin{array}{c c c c}
S_{3 1}e^{i \phi} & 0 & 0 & 0      \\
0 & S_{3 2}e^{-i \phi} & 0 & 0            \\
0 & 0 & S_{3 1}e^{i \phi} & 0  \\
0 & 0 & 0 & S_{3 2}e^{-i \phi} 
\end{array}
\right) \nonumber \\ \nonumber \\
\end{eqnarray}

\noindent where ${S_{h j}}$ are $\pm 1$ independently and depending on parameters selected. These structures were introduced in \cite{delgadoA}.

\subsubsection{Diagonal-Antidiagonal forms}

For two pulses diagonal-antidiagonal case, note that (\ref{diag}) and (\ref{adiag}) are the more general solutions to obtain an diagonal-antidiagonal evolution matrix. First and third equations in (\ref{diag2}) are still general conditions to adjust a sector on diagonal form. Second equation was used only to get a non-zero entry real, so we do not use more here. For remaining antidiagonal sector, (\ref{adiag2}) are no more the general conditions because they are specifically constructed to obtain both non zero entries just real or imaginary. Instead, (\ref{adiag}) are the general rule. Resuming, both general prescriptions give the following general diagonal ($D_2$) and antidiagonal ($A_2$)combined sectors respectively for two pulses:

\begin{eqnarray} \label{graldiagadiag}
{s_h}_j^{D_2} &=& e^{-i \alpha (J_h t+ J'_h t')} (-1)^{n_{-\alpha}} \mathbb{I}_2 \nonumber \\
&\equiv& e^{-i \alpha (J_h t+ J'_h t')} {{\mathcal S}_{h, j'}^{D_2} \mathbb{I}_2} \\
{s_h}_j^{A_2} &=& e^{i \alpha (J_h t+ J'_h t')} q_{h,j} i^h {\rm sign} (\cos {\Delta'_h}_\alpha ^- \sin {\Delta_h}_\alpha ^- {b_h}_{-\alpha} ) \cdot \nonumber \\
&& \cdot \left(
\begin{array}{c c}
0 & -e^{-i {\varphi_h}_\alpha}       \\
(-1)^h e^{i {\varphi_h}_\alpha} & 0 
\end{array}
\right) \nonumber \\
&\equiv& i e^{-i \alpha (J_h t+ J'_h t')} {{\mathcal S}_{h,j}^{A_2}}\cdot \nonumber \\
&& \cdot (\sigma_1 \sin({\varphi_h}_\alpha-\frac{\pi h}{2})+\sigma_2 \cos({\varphi_h}_\alpha-\frac{\pi h}{2}))
\end{eqnarray}

Here, we are using $-\alpha$ for diagonal sector and $\alpha$ for antidiagonal one. Script in $q_{h,j}$ should remember that
this factor is related with $\alpha$ as in (\ref{sector}). Here, $t$ and $t'$, the pulse durations, are parameters involved in the whole set of restrictions, which we repeat just for clarity. For diagonal sector $-\alpha$:

\begin{eqnarray} 
&{\Delta_h}_{-\alpha}^- + {\rm sign}({J_{\{ h \}}}_{-\alpha}({J'_{\{ h \}}}_{-\alpha}) {\Delta'_h}_{-\alpha}^- = n_{-\alpha} \pi \nonumber \\ 
&\frac{{B_h}_{\alpha}}{{J_{\{h\}}}_{-\alpha}} = \frac{{B'_h}_{\alpha}}{{J'_{\{h\}}}_{-\alpha}} \nonumber \\ 
\end{eqnarray}

\noindent with $n_{-\alpha} \in {\mathbb Z}$. This equation is easier to solve than in last section because $t$ and $t'$ are set as parameters, so this pair of equations are not coupled with those of antidiagonal sector. For antidiagonal sector $\alpha$:

\begin{eqnarray}
& {j_h}_{-\alpha}\tan {\Delta_h}_\alpha^{-}+{j'_h}_{-\alpha} \tan {\Delta'_h}_\alpha^{-} = 0 \nonumber \\ \nonumber
& {\rm sign} ({b_h}_{-\alpha}{b'_h}_{-\alpha} \sin {\Delta_h}_\alpha^- \sin{\Delta'_h}_\alpha^- \cos {\Delta_h}_\alpha^- \cos {\Delta'_h}_\alpha^-) = 1 \\ \nonumber
&{b_h}_{-\alpha}^2 \sin^2 {\Delta_h}_\alpha^- +{b'_h}_{-\alpha}^2 \sin^2 {\Delta'_h}_\alpha^-=1 \nonumber \\
\end{eqnarray}

Similarly, these set of equations can be solved by noting that second equation only adjust the correct combination of signs. Taking again $t, t'$ as parameters which lets consider ${b_h}_{\pm \alpha}$ independent from ${\Delta_h}_\alpha^-,{\Delta'_h}_\alpha^-$. Then, by expressing ${b_h}_{\pm \alpha}$ in terms of ${j_h}_{\pm \alpha}$, first and second equations can be solved simultaneously to find ${\Delta_h}_\alpha^-,{\Delta'_h}_\alpha^-$ for specific values of
${j_h}_{\pm \alpha}$. The diagonal-antidiagonal forms in two pulses become, to exchange $\left| \beta_{00} \right> \longleftrightarrow \left| \beta_{01} \right>$ or $\left| \beta_{10} \right> \longleftrightarrow \left| \beta_{11} \right>$:

\begin{eqnarray} \label{graldiagadiag1}
{U_{1}}(T) \equiv {\mathcal A}_{1, 1}^{\phi,{\varphi_h}_-} = \quad \quad \quad \quad \quad \quad \quad \quad \quad \quad \quad \quad \quad \quad \quad \quad \nonumber \\ 
=\left(
\begin{array}{c c c c}
0 & i S_{1 1}e^{i (\phi+{\varphi_h}_-)} & 0 & 0      \\
i S_{1 1}e^{i (\phi-{\varphi_h}_-)} & 0 & 0 & 0            \\
0 & 0 & S_{1 2}e^{-i \phi} & 0  \\
0 & 0 & 0 & S_{1 2}e^{-i \phi} 
\end{array}
\right) \nonumber \\
{U_{1}}(T) \equiv {\mathcal A}_{1, 2}^{\phi,{\varphi_h}_+} = \quad \quad \quad \quad \quad \quad \quad \quad \quad \quad \quad \quad \quad \quad \quad \quad \nonumber \\ 
=\left(
\begin{array}{c c c c}
S_{1 1}e^{i \phi} & 0 & 0 & 0      \\
0 & S_{1 1}e^{i \phi} & 0 & 0      \\
0 & 0 & 0 & i S_{1 2}e^{-i (\phi+{\varphi_h}_+)}  \\
0 & 0 & i S_{1 2}e^{-i (\phi-{\varphi_h}_+)} & 0 
\end{array}
\right) \nonumber \\
\nonumber \\ \nonumber \\
\end{eqnarray}

\noindent to exchange $\left| \beta_{00} \right> \longleftrightarrow \left| \beta_{11} \right>$ or $\left| \beta_{01} \right> \longleftrightarrow \left| \beta_{10} \right>$:

\begin{eqnarray} \label{graldiagadiag2}
{U_{2}}(T) \equiv {\mathcal A}_{2,1}^{\phi,{\varphi_h}_+} = \quad \quad \quad \quad \quad \quad \quad \quad \quad \quad \quad \quad \quad \quad \quad \quad \nonumber \\ 
=\left(
\begin{array}{c c c c}
0 & 0 & 0 & -S_{2 1}e^{i (\phi+{\varphi_h}_+)}      \\
0 & S_{2 2}e^{-i \phi} & 0 & 0            \\
0 & 0 & S_{2 2}e^{-i \phi} & 0  \\
s_{2 1}e^{i (\phi-{\varphi_h}_+)} & 0 & 0 & 0 
\end{array}
\right) \nonumber \\
{U_{2}}(T) \equiv {\mathcal A}_{2,2}^{\phi,{\varphi_h}_-} = \quad \quad \quad \quad \quad \quad \quad \quad \quad \quad \quad \quad \quad \quad \quad \quad \nonumber \\ 
=\left(
\begin{array}{c c c c}
S_{2 1}e^{i \phi} & 0 & 0 & 0      \\
0 & 0 & -S_{2 2}e^{-i (\phi+{\varphi_h}_-)} & 0      \\
0 & S_{2 2}e^{-i (\phi-{\varphi_h}_-)} & 0 & 0  \\
0 & 0 & 0 & S_{2 1}e^{i \phi} 
\end{array}
\right) \nonumber \\
\nonumber \\ \nonumber \\
\end{eqnarray}

\noindent to exchange $\left| \beta_{00} \right> \longleftrightarrow \left| \beta_{10} \right>$ or $\left| \beta_{01} \right> \longleftrightarrow \left| \beta_{11} \right>$:

\begin{eqnarray} \label{graldiagadiag3}
{U_{3}}(T) \equiv {\mathcal A}_{3,1}^{\phi,{\varphi_h}_-} = \quad \quad \quad \quad \quad \quad \quad \quad \quad \quad \quad \quad \quad \quad \quad \quad \nonumber \\ 
=\left(
\begin{array}{c c c c}
0 & 0 & i S_{3 1}e^{i (\phi+{\varphi_h}_-)} & 0      \\
0 & S_{3 2}e^{-i \phi} & 0 & 0            \\
i S_{3 1}e^{i (\phi-{\varphi_h}_-)} & 0 & 0 & 0  \\
0 & 0 & 0 & S_{3 2}e^{-i \phi} 
\end{array}
\right) \nonumber \\
{U_{3}}(T) \equiv {\mathcal A}_{3,2}^{\phi,{\varphi_h}_+} = \quad \quad \quad \quad \quad \quad \quad \quad \quad \quad \quad \quad \quad \quad \quad \quad \nonumber \\ 
=\left(
\begin{array}{c c c c}
S_{3 1}e^{i \phi} & 0 & 0 & 0      \\
0 & 0 & 0 & i S_{3 2}e^{-i (\phi+{\varphi_h}_+)}     \\
0 & 0 & S_{3 1}e^{i \phi} & 0  \\
0 & i S_{3 2}e^{-i (\phi-{\varphi_h}_+)} & 0 & 0 
\end{array}
\right) \nonumber \\
\nonumber \\ \nonumber \\
\end{eqnarray}

\noindent where $T = t + t'$ and second superscript, $j$, in ${\mathcal A}_{h,j}^{\phi,{\varphi_h}_\alpha}$
denotes the antidiagonal sector. An important property is that the set including ${\mathcal D}_{h}^{\phi}$ and ${\mathcal A}_{h,j}^{\phi,{\varphi_h}_\alpha}$, with ${\varphi_h}_\alpha,h,j$ fixed, forms an abelian group because ${\mathcal A}_{h,j}^{\phi',{\varphi_h}_\alpha} {\mathcal A}_{h,j}^{\phi,{\varphi_h}_\alpha}={\mathcal A}_{h,j}^{\phi,{\varphi_h}_\alpha} {\mathcal A}_{h,j}^{\phi',{\varphi_h}_\alpha}={\mathcal D}_{h}^{\phi+\phi'}$. Note that if ${\varphi_h}_\pm$ are $\frac{\pi}{2}$ and $\phi=n \pi, n \in {\mathbb Z}$ we recover forms analyzed in the previous subsection. All of them, ${\mathcal D}_h^{\phi},{\mathcal A}_{h,j}^{\phi,{\varphi_h}_\pm}$ act on Bell states to maintain them or exchange between them periodically. For other states, these operations insert programmed phases in their different components. All these operations could be written in terms of standard quantum computational gates and could give generalizations to some of these gates. Nevertheless, in the following, we will analyze only the cases with $\phi = 0, |{\varphi_h}_{\pm}|=\frac{\pi}{2}$ and will set signs ${S_{h j}}$ to have $+1$ in the first row of each sector as in (\ref{diagadiagforms}), similarly as in the standard algorithm in terms of $C^1 NOT_2$ gate.

\section{Quantum gates naturally generated by Ising interaction} 

\subsection{Equivalences with computational gates}

Diagonal-antidiagonal matrices as (\ref{diagadiagforms}) and ${\mathcal A}^{0,\pm \frac{\pi}{2}}_{h,j}$ should remember $C^a NOT_b$ gates but applied to scripts
of Bell states $\left| \beta_{\alpha \beta} \right>$ instead of those of $\left| \alpha \beta \right>$ (by extending momentarily our notation under the equivalence: $0 \leftrightarrow -, 1 \leftrightarrow +$). In fact, if we realize that ${{\mathcal A}^{0,\frac{\pi}{2}}_{1,2}}$ is almost $C^1 NOT_2 = C^1 X_2$ gate (understood as its straight form to Bell basis) except for sign changed in entry ${{\mathcal A}^{0,\frac{\pi}{2}}_{1,2}}_{+,-}$, or concisely, ${\mathcal A}^{0,\frac{\pi}{2}}_{1,2}=C^1(Z_2 X_2)$, we can get:

\begin{eqnarray}\label{equivgates}
{{\mathcal A}^{0,\frac{\pi}{2}}_{1,1}} &=& X_1 C^1 (i Y_2) X_1 \\
{{\mathcal A}^{0,\frac{\pi}{2}}_{1,2}} &=& C^1 (i Y_2) \\
{{\mathcal A}^{0,\frac{\pi}{2}}_{2,1}} &=& X_1 C^{1 \oplus 2} (i X_1 X_2) X_1 \\
{{\mathcal A}^{0,\frac{\pi}{2}}_{2,2}} &=& C^{1 \oplus 2} (i X_1 X_2) \\
{{\mathcal A}^{0,\frac{\pi}{2}}_{3,1}} &=& X_2 C^{2} (i Y_1) X_2 \\
{{\mathcal A}^{0,\frac{\pi}{2}}_{3,2}} &=& C^{2} (i Y_1)
\end{eqnarray}

\noindent where $X, Y, Z$ are Pauli matrices forms. $C^{1 \oplus 2}(G)$ is partially a classical controlled gate $G$, where control $1 \oplus 2$ means $A \oplus B$ (using the equivalence with computational scripts mentioned before) or $\frac{\alpha+1}{2} \oplus \frac{\beta+1}{2} $ (Ising notation scripts in this work), depending on form of $\left| \beta_{AB} \right> = \left| \beta_{\alpha \beta} \right>$, the scripts of state on which operator is acting. We should remember that all of these equalities are only based on matrix form, they are not the standard gates in computational basis. Anyway, this sets of operations give some alternative procedures those of quantum computing based on circuits model. 

In last sense, because the operation $({\mathcal H}_1 \otimes {\mathbb I}_2)(C^1 X_2)$
transforms computational basis into Bell basis (here, ${\mathcal H}_a$ is the Hadamard gate applied on channel $a$), whole last operations in computational basis could be expressed as $(C^1 X_2)({\mathcal H}_1 \otimes {\mathbb I}_2) ({{\mathcal A}^{\phi,{\varphi_h}_\alpha}_{h,j}}) ({\mathcal H}_1 \otimes {\mathbb I}_2)(C^1 X_2) \equiv {{{\mathcal A}^C}^{\phi,{\varphi_h}_\alpha}_{h,j}}$ and similarly for ${\mathcal D}^{\phi}_h$ ($C$ superscript denotes that gate is written in the computational basis representation). For ${{{\mathcal A}^C}^{0,\frac{\pi}{2}}_{h,j}}$ and $h = 1, 2$, its expressions shows that it transforms computational basis states into states with uniform probability distribution in that basis (with different phases inserted). For $h = 3$, this states are still exchange operators between some elements of computational basis ($\left| 00 \right> \leftrightarrow \left| 11 \right>, \left| 01 \right> \leftrightarrow \left| 10 \right>$), one at time. For ${{\mathcal D}^C}^{\phi}_h$ (defined similarly as a computational basis representation) and $h = 3$ we get an operation which introduces symmetrical phases between groups of their elements exactly as in the Bell basis representation. For $h = 1, 2$ we get an operation which generates combinations between basis elements weighted by $\sin \phi, \cos \phi$ (rotation of bipartite qubits as a whole).

\subsection{Quantum teleportation based on Ising interaction}

In addition to our interest based on Evolution loops
and Exchange operations, driven Ising operations explained
in previous section, ${\mathcal A}_{h,j}^{\phi,{\varphi_h}_\alpha}, {\mathcal D}^{\phi}_h$, suggest be used
as natural operations replacing standard gates in quantum
computing. One immediate example is teleportation
algorithm. If we begin as commonly with the state
$(\alpha \left| 0 \right>+\beta \left| 1 \right>) \otimes \left| \beta_{--} \right>$ with the first qubit to teleport in possession of Alice and Bell state first in possession of Bob, who send its first part to Alice. Alice drives Ising interaction as in our model applying ${\mathcal A}_{1,2}^{0,\frac{\pi}{2}}$ operation on them. With this, they almost obtain the standard teleportation algorithm for one qubit \cite{bennet4}:

\begin{eqnarray}\label{teleportation}
&& (\alpha \left| 0 \right> + \beta \left| 1 \right>) \otimes \left| \beta_{00} \right> = \nonumber \\
&& \quad \quad \frac{\alpha}{2} \bigg( (\left| \beta_{00} \right>+\left| \beta_{10} \right>) \otimes \left| 0 \right> + 
   (\left| \beta_{01} \right>+\left| \beta_{11} \right>) \otimes \left| 1 \right> \bigg) \nonumber \\
&& \quad + \frac{\beta}{2} \bigg( (\left| \beta_{01} \right>-\left| \beta_{11} \right>) \otimes \left| 0 \right> 
	 (\left| \beta_{00} \right>-\left| \beta_{10} \right>) \otimes \left| 1 \right> \bigg) \nonumber \\
&& \stackrel{{\mathcal A}_{1,2}^{0,\frac{\pi}{2}}}{\longrightarrow} \frac{\alpha}{2} \bigg( (\left| \beta_{00} \right>-\left| \beta_{11} \right>) \otimes \left| 0 \right> + 
(\left| \beta_{01} \right>+\left| \beta_{10} \right>) \otimes \left| 1 \right> \bigg) \nonumber \\
&& \quad + \frac{\beta}{2} \bigg( (\left| \beta_{01} \right>-\left| \beta_{10} \right>) \otimes \left| 0 \right> + 
(\left| \beta_{00} \right>+\left| \beta_{11} \right>) \otimes \left| 1 \right> \bigg) \nonumber \\ \nonumber \\
\end{eqnarray}

\noindent where for simplicity, in order to identify easily the results in computational basis, we are comeback briefly to classical notation for Bell states. 

After to apply Ising interaction, Alice makes a measurement
in computational basis on two first qubits. Results
are listed in Table I where $\left| - \right>, \left| + \right>$ are the eigenstates of $X$. Thus, Alice uses classical communication to send outcomes to Bob who finally apply adequate gates (Table I) to obtain the original state. It is clear that Bob will require apply a Hadamard gate at the end and if measurement result is $\left| AB \right>$, then it is precise to apply in addition $X^B Z^{1+A}$ to teleportate the original qubit on qubit 3 in his possesion. Another alternative for Alice is make a measurement in Bell basis $\left| \beta_{\alpha \beta} \right>$, then to send the result to Bob, in such case, he will need apply $Z^{\frac{1+\alpha}{2}} X^{\frac{1-\alpha \beta}{2}}$. 

\begin{table}[htb]
 \centering \caption{Measurement outcomes for qubits $1$ and $2$, output state in qubit $3$ and complementary gates for teleportation algorithm based on ${\mathcal A}_{1,2}^{0,\frac{\pi}{2}}$. First four rows after of titles corresponds to measurements in the computational basis and last four to Bell basis measurement.}
\begin{tabular}{|c| c | c|}
    \hline
    Measurement & Teleportated state & Complementary gates \\
    \hline
		\hline
    $\left|00 \right>$ & $ \quad \alpha \left|+ \right>-\beta \left|- \right>$ & $Z_3{\mathcal H}_3$ \\
    \hline
		$\left|01 \right>$ & $ -\alpha \left|- \right>+\beta \left|+ \right>$ & $X_3Z_3{\mathcal H}_3$ \\
    \hline
		$\left|10 \right>$ & $ \quad \alpha \left|+ \right>+\beta \left|- \right>$ & ${\mathcal H}_3$ \\
    \hline
		$\left|11 \right>$ & $ \quad \alpha \left|- \right>+\beta \left|+ \right>$ & $X_3{\mathcal H}_3$ \\
    \hline
		\hline
		$\left|\beta_{--} \right>$ & $ \quad \alpha \left|0 \right>+\beta \left|1 \right>$ & ${\mathcal I}_3$ \\
    \hline
		$\left|\beta_{-+} \right>$ & $ \quad \alpha \left|1 \right>+\beta \left|0 \right>$ & $X_3$ \\
    \hline
		$\left|\beta_{+-} \right>$ & $ \quad \alpha \left|1 \right>-\beta \left|0 \right>$ & $Z_3 X_3$ \\
    \hline
		$\left|\beta_{++} \right>$ & $ -\alpha \left|0 \right>+\beta \left|1 \right>$ & $Z_3$ \\
    \hline
   \end{tabular}
    \end{table}
		
\noindent where ${\mathcal I}$ is the identity gate for a single qubit. This example shows that physical interactions, as driven Ising model, could be used to reproduce standard gates proposed in quantum computing, or otherwise, those gates can be equivalently replaced by quantum physical processes.

\section{Entanglement dynamics} \label{secvi}

\subsection{Evolution of Bell states entanglement}

As was obtained in \cite{delgadoA}, concurrence of states evolved from Bell states, $\left|\beta_{\mu \nu} \right>$, with $U_h(t)$ from (\ref{hamiltonian}) becomes:

\begin{eqnarray}\label{bellentanglement}
{\mathcal C}^h_{\mu \nu} &=& 1-4 {j_h}_{-f^h_{\mu \nu}}{b_h}_{-f^h_{\mu \nu}} \sin^4 {\Delta_h^-}_{f^h_{\mu \nu}} \\
f^h_{\mu \nu} &=& 
\left\{
\begin{array}{cc}
\mu & , h=1 \\ 
\mu \nu & , h=2 \\  
\nu & , h=3  
\end{array} \right. \nonumber
\end{eqnarray}

\noindent exhibiting a simple behavior depending only on one Rabi frequency, in contrast with evolution of other states, specially separable ones. This expression is consistent with isotropic case reported in \cite{delgado1,delgado2} and with results (\ref{evloop2}) and (\ref{diagadiag2}-\ref{diagadiag3}). But it implies a more general feature in spite that not for all physical parameters in the current interaction, Bell states comeback each period to their original states, but instead into another state with maximal entanglement, denoting a natural correspondence between maximal entanglement with current Ising interaction model. In another perspective, intermediate stages of evolution for initially Bell states does not always become separable. Only if ${j_h}^2_{f^h_{\mu \nu}} {b_h}^2_{f^h_{\mu \nu}}=1/4$, then ${\mathcal C}^h_{\mu \nu}$ reaches its maximum value. It means, when $|{B_h}_{-f^h_{\mu \nu}}|=|{J_{\{h\}}}_{f^h_{\mu \nu}}|$, the perfect tuning between field and strength interaction. 

\subsection{Control of maximal entanglement} 

As in \cite{delgado2}, presence of parasitic magnetic fields could affect the stability of maximal entanglement. Thus, if qubits are currently exposed to magnetic fields ${B_1}_h,{B_2}_h$ respectively, then maximal entangled values of evolution of Bell states have oscillations with frequencies $\omega_- =
{R_h}_-, \omega_+ = {R_h}_+$ by pairs in agreement with (\ref{erres}) and (\ref{bellentanglement}).
In that situation, an homogeneous magnetic field, ${B_h}_0$,
can be added to tune the oscillation frequency of entanglement of all Bell states under evolution by requiring: $n_+ {R'_h}_- = n_- {R'_h}_+$ (where $n_+, n_-\in {\mathbb Z}^+$). Magnetic field ${B_h}_0$ becomes:

\begin{equation}
{B_h}_0=\frac{1}{2}(-{B_h}_+ \pm n_+ \sqrt{{R_h}_-^2-{J_{\{ h \}}}_+^2})
\end{equation}

Other possible desired effect is maximize the oscillation
amplitude for concurrence in order to assure that there
are periodical intermediate separable states. In this case,
an homogeneous field ${B_h}_0$ is sufficient to reach this effect for those Bell states associated in (\ref{bellentanglement}) with ${B_h}_+$ for each value of $h$. We need to get a new effective field $|{B'_h}_+|=|{J_{\{h\}}}_-|$, it implies:

\begin{equation}
{B_h}_0=\frac{1}{2}(-{B_h}_+ \pm |{J_{\{ h \}}}_-|)
\end{equation}

Nevertheless, for Bell states associated with ${B_h}_-$ in
(\ref{bellentanglement}), selective fields on each qubit $i = 1, 2$, ${B'_i}_h = {B_i}_h +\delta {B_i}_h$, should be applied, in such way that they fulfill:

\begin{equation}
\delta {B_1}_h - \delta {B_2}_h=-{B_h}_- \pm |{J_{\{ h \}}}_+|)
\end{equation}

Finally, it is clear that under these kind of effects, it is possible nullify
${B_h}_-$ or ${B_h}_+$ in order to set invariant some Bell states (as in \cite{delgado1} for isotropic Ising interaction). In this cases, some of eigenvalues (\ref{eigenvectors}) become Bell states.

\section{Conclusions}

Nowadays, research about physical systems on which set up quantum technology, in particular that related with quantum computation, quantum information processing or quantum cryptography are growing since several directions to get them in an useful form for those applications. Nevertheless that optics has been partially a dominant arena to last developments, matter has been shown several benefits. Quantum storage and processing of information allow new computational tasks which are impossible with conventional information technology or quantum optics exclusively. In this trend, systems based on trapped ions, e-Helium, nuclear magnetic resonance, superconductors, doped silicon and quantum dots have shown opportunities to make stable and efficient developments for that purpose \cite{klo1}.

Such quantum stuff requires a system of several qubits. The main materials based technology known for that purpose is magnetic. For this reason, spin-based quantum computing has been developed in several experimental implementations, which uses magnetic systems mainly: superconducting integrated systems, superconducting flux qubits, straight nuclear magnetic resonance and quantum dots. All of them exploits Ising interactions with different approaches \cite{john1}, together with control on quantum states and in particular with entanglement control, a milestone in all almost these researches.

Circuit-gate model was the first approach to quantum computation, nevertheless, quantum annealing \cite{kado1} or measurement-based quantum computation \cite{briegel1} are alternatives which use magnetic systems approached by Ising interactions to manage a planned and controlled quantum state manipulation. On them, several applied problems has been exhibited as the goal of (these technologies (pattern matching, folding proteins, an other particular NP-complete problems \cite{john1}).

Several aspects around of stability to set these elements as isolated qubits has been solved in parallel with their control of quantum properties as those here studied. Nevertheless than in the current time, deep control of parameters in quantum magnetic systems is still limited, the study of Ising model in a comprehensive way, by including several freedom degrees as manipulable (in particular Ising interaction strengths) opens future opportunities to explode all their computable possibilities. Otherwise, control of time and magnetic fields is very well developed. \cite{klo1} reports experimental data about magnetic control parameters for these systems in related problems to presented here. They are located around of $t \sim 10^{-9}-10^{-6} s$ and $B \sim 1-10 T$ settled in regions with sizes around of $r \lesssim 5 nm$, which are values completely achievable in the contemporary control physics.

Still, a programmable artificial spin network should be constructed on bipartite qubits behavior knowledge \cite{john1}. In this sense, current work is developing this basic knowledge with models with additional freedom degrees. Basic form of evolution in (\ref{mathamiltonian}) have a regular structure in terms of group theory \cite{delgadoA}. Gates are constructed in this model as physical operations to reproduce a planned evolution to simulate some problem based on quantum resources being used. Finally, those operations should be constructed on basic controlled operations similar to Evolution Loops or Exchange Operations as presented here, as Evolution loops achieved exactly as ${\mathbb I}_4$ or exchange operations achieved with forms (ref{diagadiagforms}). These operations are useful by themselves in discrimination and/or quantum state correction \cite{delgado2}.

Still, these operations, more than their own properties, are settled on an specific basis: Bell states. By combining these operations (\ref{mathamiltonian}) in different directions we can increase possibilities about entanglement control as stated in Figure \ref{fig1} or in terms of basic results in section \ref{secvi}. The creation of universal procedures to reproduce arbitrary gates is open in terms of elements identified as controlled operations (\ref{equivgates}). 

Gates to reproduce well known procedures in circuit-gate quantum computation (as teleportation, quantum Fourier transform, etc.) should be adapted each time in function of quantum system which has been used as implementation, then current model unifies magnetic systems in order to reproduce those gates. Possible adaptability in quantum computation for this purpose can generate a universal, or at least, a easier way of design.

In this line of research, the analysis of behavior with finite temperature based on matrix density is required to consider realistic decoherence effects. In addition, error correction analysis is necessary in our procedures, based on experimental limitations factors (control on time and magnetic field, knowledge and control of interaction strengths, etc.). In our approach, an improvement should be generated through alternative continuous pulses (by example $B(t)=B_0+B_p \sin (\omega t)$, instead of rectangular as used here. Rectangular pulses are easy to manage theoretically but they are experimentally few practical because of their discontinuity and their associated resonant effects.

Nuclear magnetic resonance, Quantum dots and electrons in silicon lattices have been the most successful systems in implementing quantum algorithms based on their coherence and stability \cite{klo1}. These systems could explode two aspects presented theoretically here, first is try to state non local basis as a natural language because they appear as more natural in the interactions involved (despite of course of their own difficulties to avoid decoherence in this states). In addition, traduce classical gates in circuit-gate model of quantum computation into alternative gates as present in this work, nevertheless that they could do not appear as direct in terms of classical bit manipulation.

\small  

\end{document}